\documentclass{article}
\usepackage{spconf,amsmath,graphicx}
\usepackage{tabularx}
\usepackage{color}
\usepackage{booktabs}
\usepackage{multirow}
\usepackage{tikz}
\usepackage{cite}
\usetikzlibrary{calc,fit,positioning, backgrounds}
\usetikzlibrary{external}
\usepackage{graphbox}

\colorlet{dark green}{green!70!black}
\colorlet{light blue}{blue!30!white}
\colorlet{dark red}{red!50!black}
\newlength{\distance}
\newlength{\plotheight}
\title{Experimental Investigation on STFT Phase Representations for \\ Deep Learning-Based Dysarthric Speech Detection}
\name{Parvaneh Janbakhshi$^{*,\dagger}$, Ina Kodrasi$^{*}$\thanks{This work was supported by the Swiss National Science Foundation, project no CRSII5\_173711 on ``Motor Speech Disorders: characterizing phonetic speech planning and motor speech programming/execution and their impairments'' and project no CRSII5\_202228 on ``Characterisation of motor speech disorders and processes''.}}
\address{$^{*}$Idiap Research Institute, Martigny, Switzerland \\
  $^{\dagger}$\'Ecole Polytechnique F\'ed\'erale de Lausanne, Lausanne, Switzerland\\
  \tt \{parvaneh.janbakhshi,ina.kodrasi\}@idiap.ch\\
}
\newcommand{\bftab}{\fontseries{b}\selectfont}

\begin{document}
\ninept
\maketitle
\begin{abstract}
  Mainstream deep learning-based dysarthric speech detection approaches typically rely on processing the magnitude spectrum of the short-time Fourier transform of input signals, while ignoring the phase spectrum.
  Although considerable insight about the structure of a signal can be obtained from the magnitude spectrum, the phase spectrum also contains inherent structures which are not immediately apparent due to phase discontinuity.
  To reveal meaningful phase structures, alternative phase representations such as the modified group delay~(MGD) and instantaneous frequency~(IF) spectra have been investigated in several applications.
  The objective of this paper is to investigate the applicability of the unprocessed phase, MGD, and IF spectra for dysarthric speech detection.  
  Experimental results show that dysarthric cues are present in all considered phase representations.
  Further, it is shown that using phase representations as complementary features to the magnitude spectrum is beneficial for deep learning-based dysarthric speech detection, with the combination of magnitude and IF spectra yielding a high performance.
  The presented results should raise awareness in the research community about the potential of the phase spectrum for dysarthric speech detection and motivate research into novel architectures which optimally exploit magnitude and phase information.
\end{abstract}
\section{Introduction}
\label{sec: intro}
  Dysarthria is a motor speech disorder arising from different conditions of brain damage and manifesting through articulation deficiencies, vowel quality changes, abnormal speech rhythm, pitch variation, or breathiness~\cite{Duffy_book_2003}.
  Since dysarthria can be one of the earliest signs of several neurodegenerative disorders, its accurate diagnosis in clinical practice is crucial~\cite{Rusz_JASA_2013, tracy_bi_2020}.
  The clinical diagnosis of dysarthria is typically done through an auditory-perceptual approach, which can be subjective and time-consuming.
  To complement the clinical perceptual assessment, automatic dysarthric speech detection approaches have been developed.
  
  Automatic dysarthric speech detection approaches can be broadly categorized into two categories, i.e., i) approaches based on handcrafted acoustic features combined with classical machine learning algorithms~\cite{Tsanas_ITBE_2012, Orozco-Arroyave_Interspeech_2015, Kodrasi_ITASLP_2019a, Janbakhshi_SPL_2020, Hernandez_IS_2020} and ii) deep learning-based approaches that automatically learn high-level discriminative dysarthric representations~\cite{Janbakhshi_ICASSP_2021, Vasquez2017, Vaiciukyna_SOTSG_2017, Kwanghoon_IS_2018, Vasquez_SC_2020, Janbakhshi_ITG_2021}.
  Given the potential of deep learning-based approaches to characterize abstract but important acoustic cues beyond the realm of knowledge-based handcrafted features, in this paper we focus on deep learning-based approaches.

  Mainstream deep learning-based dysarthric speech detection approaches rely on processing the magnitude spectrum (or features derived from the magnitude spectrum) of time-frequency representations such as the short-time Fourier transform~(STFT) or continuous wavelet transform~\cite{Vasquez2017, Vaiciukyna_SOTSG_2017, Kwanghoon_IS_2018, Vasquez_SC_2020, Janbakhshi_ITG_2021}.
  In~\cite{Vasquez2017}, articulation impairments of patients suffering from dysarthria are modeled through a convolutional neural network~(CNN) operating on the magnitude spectrum of the continuous wavelet transform.
  In~\cite{Vaiciukyna_SOTSG_2017, Kwanghoon_IS_2018}, a CNN is trained on the STFT magnitude spectrum or Mel frequency cepstral coefficients of neurotypical and dysarthric input signals.
  The STFT magnitude spectrum is also used in~\cite{Vasquez_SC_2020, Janbakhshi_ITG_2021} to train unsupervised and supervised auto-encoders for dysarthric speech detection.
  Although considerable insight about the structure of a signal can be obtained from the magnitude spectrum, there are inherent structures also in the phase spectrum, which however has been largely ignored in automatic dysarthric speech detection techniques.

  The disregard of the phase spectrum in speech processing applications arises mainly due to the difficulty in processing the phase and due to the uncertainty about its importance~\cite{Mowlaee_INTERSPEECH_2014,Gerkmann_SPLmag_2015}.
  Since phase is wrapped to its principal value, the phase spectrum is discontinuous.
  Consequently, the phase spectrum is irregular and does not contain visible spectro-temporal patterns that correlate with our understanding of speech.
  However, several methods have been developed to derive alternative representations revealing spectro-temporal structures hidden in the phase spectrum.
  Two such representations are the modified group delay~(MGD) spectrum~\cite{Yegnanarayana_ICSLP_1990,Murthy_ICASSP_2003} and the instantaneous frequency~(IF) spectrum~\cite{Friedman_ICASSP_1985,Boashash_IEEE_1992}.
  The MGD and IF spectra reflect the derivative of phase along the frequency and time axis and have been shown to reveal much more meaningful structures than the unprocessed phase spectrum~\cite{Boashash_IEEE_1992}.
  In addition, although early studies have demonstrated the unimportance of phase to speech perception~\cite{Oppenheim_IEEE_1981, Wang_ITASLP_1982}, more recent studies have established the potential of the phase spectrum in different applications such as speech enhancement~\cite{Gerkmann_SPL_2013, Gerkmann_SPLmag_2015}, automatic speech recognition for neurotypical and dysarthric speech~\cite{Hedde_ITASLP_2007,Sehgal_STL_2018}, or speech synthesis~\cite{Degottex_ITASLP_2011}.
  The potential of the phase spectrum has also been demonstrated for computational paralinguistic applications such as speaker recognition~\cite{Nakagawa_ITASLP_2012, Rajan_INTERSPEECH_2013} and speech emotion recognition~\cite{Deng_IEEEAcess_2016, Guo2018}.

  To the best of our knowledge, the STFT phase spectrum or its alternative representations such as the MGD or IF spectra have never been incorporated in deep learning-based dysarthric speech detection approaches.
  As shown in~\cite{Yegnanarayana_ICSLP_1990,Murthy_ICASSP_2003}, the MGD spectrum reflects the formant structure of the speech signal.
  Further, as shown in~\cite{Friedman_ICASSP_1985}, the IF spectrum for narrow-band analysis displays pitch information in the form of fine harmonic detail.
  Since, the MGD spectrum can be expected to capture articulation deficiencies and vowel quality changes and the IF spectrum can be expected to capture pitch variation, we expect such representations to be useful for dysarthric speech detection.

  In a recently proposed approach, we have used the temporal envelope and fine structure (i.e., analytical phase) representations of speech signals computed through a Gammatone filter bank and sub-sampling~\cite{Kodrasi_SPL_2021}.
  These input representations are separately processed by CNNs to learn two discriminative representations, which are then jointly used for dysarthric speech detection.  
  Experimental results in~\cite{Kodrasi_SPL_2021} show that such an approach yields a considerable performance improvement when compared to processing only the STFT magnitude spectrum.
  However, it remains unclear whether this substantial performance increase arises because of the incorporation of the analytical phase or because of the auditory-inspired processing through a Gammatone filter bank instead of a uniform STFT filter bank.
  
  In this paper we investigate the applicability of STFT phase representations (i.e., the unprocessed phase spectrum, MGD spectrum, and IF spectrum) for dysarthric speech detection.
  To this end, we analyze the dysarthric speech detection performance of a CNN which uses only phase representations of input signals.
  In addition, we analyze whether phase representations provide complementary cues for dysarthric speech detection that cannot be extracted from the magnitude representation.
  Experimental results show that dysarthric cues are present in all considered phase representations, with the MGD and IF spectra yielding a similar dysarthric speech detection performance as the magnitude spectrum.
  Further, it is shown that using both the magnitude and any of the phase representations is beneficial, resulting in a considerable performance improvement as opposed to using a single representation.
  Finally, it is shown that using the STFT magnitude and IF spectra results in a considerably better performance than using the temporal envelope and fine structure representations from~\cite{Kodrasi_SPL_2021}.

\section{Input Representations}
\label{sec: allreps}
In this section, the computation of input signal representations is presented.
Section~\ref{sec: reps} presents the STFT phase representations investigated in this paper.
For completeness, Section~\ref{sec: tefs} provides a brief review of the temporal envelope and fine structure representations used in~\cite{Kodrasi_SPL_2021}.

\subsection{STFT magnitude and phase representations}
\label{sec: reps}
\begin{figure}[t]
  \centering
  \includegraphics[scale=0.55]{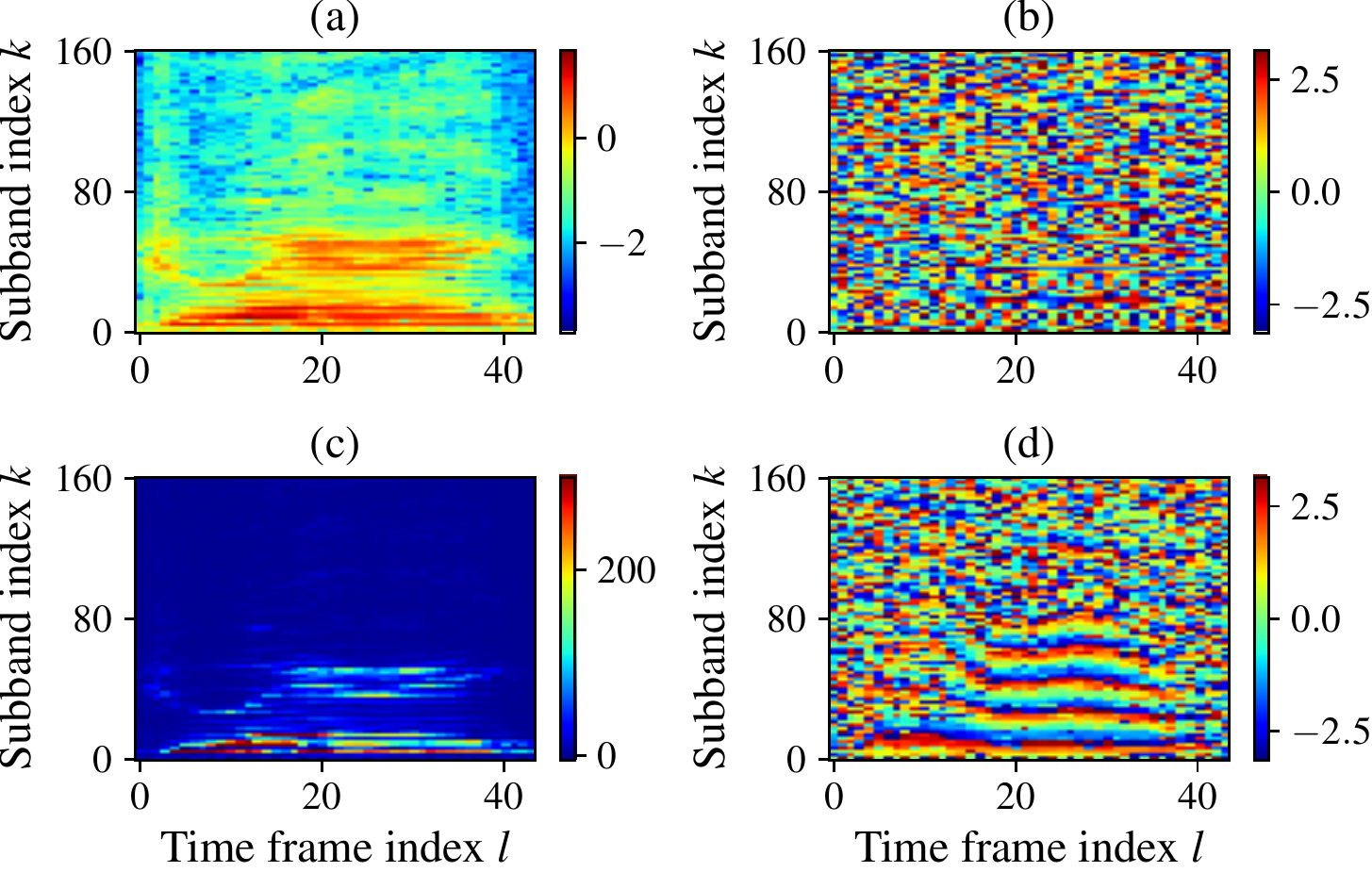}
  \caption{STFT representations of an exemplary utterance computed using $N=320$ samples with a $50$\% overlap and a Hanning analysis window: (a) logarithm of the magnitude, (b) phase, (c) modified group delay, and (d) instantaneous frequency.}
  \label{fig: transform}
\end{figure}

Let us consider the time-domain signal $s(n)$ at time index $n$.
To compute the STFT, the signal is segmented into $L$ segments $s_l(n)$ (with or without overlap) of length $N$ samples.
Each segment $s_l(n)$ is multiplied by an analysis window and the discrete Fourier transform is applied to obtain the complex STFT coefficients $S_{k,l}$, $k = 1, \ldots, K,$ with $K$ denoting the total number of subbands.
The complex STFT coefficients can be expressed as
\begin{equation}
  S_{k,l} = |S_{k,l}| e^{j \theta_{k,l}},
\end{equation}
with $|S_{k,l}|$ and $\theta_{k,l}$ denoting the magnitude and phase of the $l$-th segment at the $k$-th subband.
Figs.~\ref{fig: transform}(a) and~\ref{fig: transform}(b) depict the magnitude and phase spectra for an exemplary utterance $s(n)$.
It can be observed that while the magnitude spectrum exhibits spectro-temporal patterns where formant and pitch information can be identified, the phase spectrum is irregular and difficult to interpret since the phase is wrapped to its principal value, i.e., $-\pi \leq \theta_{k,l} \leq \pi$.

In this paper we investigate the applicability of two alternative phase representations in deep learning-based dysarthric speech detection which aim to reveal spectro-temporal structures hidden in the phase spectrum, i.e., the modified group delay~(MGD) spectrum~\cite{Yegnanarayana_ICSLP_1990,Murthy_ICASSP_2003} and the instantaneous frequency~(IF) spectrum~\cite{Friedman_ICASSP_1985,Boashash_IEEE_1992}.

The group delay is defined as the negative of the derivative of phase across frequency and can be computed as
\begin{equation}
  \label{eq: gd}
  \tau_{k,l} = \frac{S^{\rm r}_{k,l} Y^{\rm r}_{k,l} + S^{\rm i}_{k,l} Y^{\rm i}_{k,l}}{|S_{k,l}|^2},
\end{equation}
with $Y^{\rm r}_{k,l}$ and $Y_{k,l}^{\rm i}$ denoting the real and imaginary part of the STFT coefficients $Y_{k,l}$ of $y_l(n) = ns_l(n)$.
To reduce the spiky nature of the group delay spectrum in~(\ref{eq: gd}), the MGD spectrum is proposed in~\cite{Murthy_ICASSP_2003}, which can be computed as
\begin{equation}
  \label{eq: mgd}
{\rm MGD}_{k,l} = {\rm sign} \left\{ \frac{S^{\rm r}_{k,l} Y^{\rm r}_{k,l} + S^{\rm i}_{k,l} Y^{\rm i}_{k,l}}{\hat{S}_{k,l}^{2\gamma}} \right\}  \left| \frac{S^{\rm r}_{k,l} Y^{\rm r}_{k,l} + S^{\rm i}_{k,l} Y^{\rm i}_{k,l}}{\hat{S}_{k,l}^{2\gamma}} \right|^{\alpha}\!\!\!\!,
\end{equation}
where $\hat{S}_{k,l}$ denotes the cepstrally smoothed version of $|S_{k,l}|$ and $\alpha$ and $\gamma$ are hyper-parameters controlling the spiky nature of the resulting spectrum~\cite{Murthy_ICASSP_2003}.

The IF spectrum is defined as the derivative of phase across time and can be computed as~\cite{Kay_ITASSP_1989, Stark_Interspeech_2008}
\begin{equation}
  \label{eq: if}
  {\rm IF}_{k,l} = \arg \{S_{k,l+1} S^{*}_{k,l} \},
\end{equation}
with $\arg \{ \cdot \}$ denoting the complex phase function and $\{\cdot \}^{*}$ denoting the complex conjugate.
Using~(\ref{eq: if}) to compute the IF spectrum helps to partly alleviate phase wrapping issues~\cite{Kay_ITASSP_1989, Stark_Interspeech_2008}.

Figs.~\ref{fig: transform}(c) and~\ref{fig: transform}(d) depict the MGD and IF spectra for the previously considered exemplary utterance $s(n)$, with magnitude and phase spectra depicted in Figs.~\ref{fig: transform}(a) and~\ref{fig: transform}(b).
It can be observed that contrary to the phase spectrum and similarly to the magnitude spectrum, both the MGD and IF spectra exhibit regular spectro-temporal patterns which can be potentially exploited by deep learning-based dysarthric speech detection approaches.

\subsection{Temporal envelope and fine structure representations}
\label{sec: tefs}
Instead of computing signal representations based on the uniform STFT filter bank, in~\cite{Kodrasi_SPL_2021} we have proposed to compute the temporal envelope and fine structure representation using Gammatone filter banks mimicking cochlear frequency analysis.
To this end, the signal $s(n)$ is split into $K$ complementary frequency bands of equal width along the human basiliar membrane~\cite{Smith_nature_2002}.
Let us denote by $s^c_{k}(n)$ the signal obtained at the output of the $k$-th band pass filter.
The analytic representation of $s^c_{k}(n)$ is given by
\begin{equation}
  \label{eq: an}
  s^a_{k}(n) = s^c_{k}(n) + j {\cal{H}} \{s^c_{k}(n) \},
\end{equation}
where ${\cal{H}} \{ \cdot \}$ denotes the Hilbert transform.
The magnitude and cosine of the phase of the complex coefficients in~(\ref{eq: an}) yield the temporal envelope and fine structure signals.
These signals are sub-sampled by taking the mean over sliding windows of length $N$ samples (with or without overlap) to obtain the final temporal envelope and fine structure representations used for dysarthric speech detection in~\cite{Kodrasi_SPL_2021}.

\section{Magnitude and Phase-based \\ Dysarthric Speech Detection}
\label{sec: networks}
\begin{figure}[t!]
  \hspace{1.8cm} (a) \hspace{4cm} (b)
  
  \includegraphics[align=c,scale=0.85]{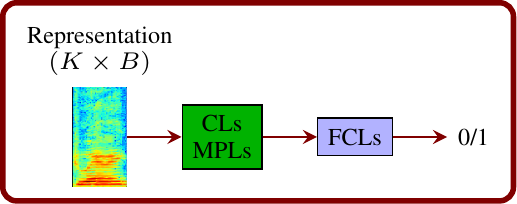}
  \includegraphics[align=c,scale=0.85]{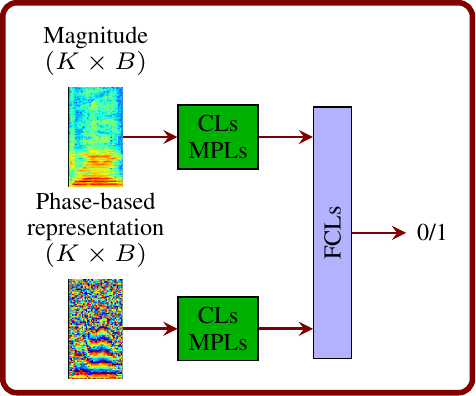}
  \caption{Block diagram of the considered CNN-based dysarthric speech detection approach: (a) single input approach operating on the magnitude, phase, MGD, or IF spectrum and (b) dual input approach operating on the magnitude and phase spectra, magnitude and MGD spectra, or magnitude and IF spectra. CLs, MPLs, and FCLs refer to convolutional, max-pooling, and fully-connected layers, respectively.}
  \label{fig: scheme_seg}
\end{figure}

To investigate the applicability of phase representations for dysarthric speech detection, we consider the state-of-the-art CNN-based approach from~\cite{Vasquez2017} depicted in Fig.~\ref{fig: scheme_seg}(a).
As shown in this figure, the CNN operates on $(K \times B)$--dimensional magnitude representations, with $B$ denoting a user-defined number of time frames~(cf. Section~\ref{sec: results}).
Through alternating convolutional and max-pooling layers, the network learns a discriminative representation from the magnitude spectrum of neurotypical and dysarthric signals.
In Section~\ref{sec: results} we investigate the performance of this approach when $(K \times B)$--dimensional phase representations (i.e., unprocessed phase, MGD, IF) are used as input instead of the magnitude spectrum used in~\cite{Vasquez2017}.

To further analyze whether phase representations provide additional cues that cannot be extracted from the magnitude, we consider the dual input CNN from~\cite{Kodrasi_SPL_2021} depicted in Fig.~\ref{fig: scheme_seg}(b).
In~\cite{Kodrasi_SPL_2021}, this dual input CNN operates on $(K \times B)$--dimensional envelope and fine structure representations computed as described in Section~\ref{sec: tefs}.
As shown in this figure, different convolutional and max-pooling layers are used on the different input representations.
Hence, two different discriminative representations are learned and jointly exploited through fully-connected layers to detect dysarthric speech.
Instead of using the temporal envelope and fine structure representations, in this paper we investigate the performance of the dual input CNN operating on the STFT magnitude spectrum and phase representations, i.e., magnitude and phase spectra, magnitude and MGD spectra, or magnitude and IF spectra.

\section{Experimental Results}
\label{sec: results}
In this section, the dysarthric speech detection performance of the single and dual input CNNs is compared for different input representations.
A PyTorch implementation of the considered approaches is available online.\footnote{https://github.com/idiap/pddetection-phase-reps}

\subsection{Database}
\label{sec: database}
We consider recordings of $24$ different words and a phonetically balanced text from $50$ neurotypical speakers and $50$ PD patients from the well-balanced PC-GITA database~\cite{GITA}.
The recordings are downsampled to $16$~kHz from the original sampling frequency of $44.1$~kHz.
The average length of the total speech material available for each speaker is $32.1$ seconds.

\subsection{Input representations and network architectures}
\label{sec: compreps}
The STFT is obtained using a weighted overlap-add framework with a Hanning analysis window without overlap and a frame size of $N = 160$ samples (i.e., $10$~ms), resulting in $K = 81$ subbands.
The logarithm of the magnitude spectrum, the phase spectrum, and the IF spectrum are computed from the STFT coefficients~(cf.~Section~\ref{sec: reps}).
To compute the MGD spectrum, we use a cepstral window of length $20$ samples, $\alpha = 0.6$, and $\gamma = 0.3$~(cf.~(\ref{eq: mgd})).
The temporal envelope and fine structure representations are also computed using $10$~ms segments without overlap and $K=81$ auditory filter banks.
The remainder of the parameters used in computing the temporal envelope and fine structure representations are the same as in~\cite{Kodrasi_SPL_2021}.
Similarly to~\cite{Kodrasi_SPL_2021}, ($K \times B$)--dimensional segments using $B = 50$ and a $50$\% overlap are extracted from the computed representations and used as inputs to the CNNs.
Input representations are normalized to a mean of $0$ and a standard deviation of $1$.

The architecture of the single input CNN in Fig.~\ref{fig: scheme_seg}(a) consists of two convolutional layers with $64$ channels, a $2\times2$ kernel for the first layer, and a $3 \times 3$ kernel for the second layer.
Each convolutional layer is followed by a ReLU activation function, batch normalization, and max-pooling with a $2 \times 2$ kernel.
The second convolutional layer is followed by a dropout layer with a rate of $0.5$.
After the dropout layer, a fully-connected layer (input dimension of $4224$ and output dimension of $2$) followed by the softmax function is used.

The dual input CNN in Fig.~\ref{fig: scheme_seg}(b) has the same architecture of convolutional, max-pooling, and dropout layers for the upper and lower branches as the single input CNN.
The output of these two branches is fused through a fully-connected layer with an input size of $8448$, an output size of $128$, and a ReLU activation function.
A final fully-connected layer with an input size of $128$ and an output size of $2$ followed by the softmax function is then used.

\subsection{Training and evaluation}
\label{sec: training}
The performance of the considered approaches is evaluated in a speaker-independent stratified $10$-fold cross-validation framework.
The stochastic gradient descent algorithm and cross-entropy loss are used for training.
The batch size is $128$ and the initial learning rate is $0.01$.
In each training fold, a development set with the same size as the test set is used, such that the learning rate is halved if the loss on the development set does not decrease for $5$ consecutive iterations.
Training is stopped when the learning rate has decreased beyond $10^{-6}$ or after $100$ epochs.
The single input CNNs are randomly initialized. 
The convolutional layers of the dual input CNNs are initialized with the convolutional layers of the trained single input counterparts. 

The final prediction score for a test speaker is obtained through soft voting of the prediction scores obtained for each $(K\times B)$--dimensional input representation belonging to the speaker.
Dysarthric speech detection performance is evaluated in terms of the area under ROC curve (AUC) and classification accuracy for a decision threshold of $0.5$.
To reduce the impact that the random initialization of networks and the random split of speakers into training and testing folds have on the final performance, we have trained all networks with $5$ different random seeds for $5$ different splits of speakers.
The reported performance measures are the mean and standard deviation of the performance obtained across these different models.

\subsection{Results}
\label{sec: results}

To investigate the applicability of phase representations in comparison to the traditionally used magnitude spectrum, we analyze the performance of the single input CNN operating on different input representations.
Table~\ref{tbl: perf_single} presents the performance of the single input CNN operating on the magnitude, phase, MGD, and IF spectra.
It can be observed that using the IF spectrum yields the highest performance in terms of accuracy, with an AUC score similar to the AUC score obtained when using the MGD spectrum.
Further, it can be observed that the performance when using the magnitude and MGD spectra is very similar.
This result is to be expected since as it can be visually inspected in Fig.~\ref{fig: transform}, the MGD spectrum contains spectro-temporal patterns that are similar to the magnitude spectrum.
The lowest performance in terms of accuracy and AUC score is obtained when using the phase spectrum, which is also to be expected since the phase spectrum is irregular and visually void of meaningful structures.
However, it should be noted that using the phase spectrum yields an accuracy of $62.76$\% and an AUC score of $0.70$, which shows that although the phase spectrum does not visually exhibit any regular spectro-temporal structures, a CNN nevertheless manages to partially discover cues in the phase spectrum that are important for dysarthric speech detection.

\begin{table}[t]
  \footnotesize
  \begin{center}
    \caption{\footnotesize Performance of the single input CNN operating on magnitude and phase representations.}
    \label{tbl: perf_single}
    \begin{tabularx}{\linewidth}{Xrr}
      \toprule
      Representation & Accuracy & AUC \\
      \toprule
      Magnitude & 69.72 $\pm$ 15.62 & 0.77 $\pm$ 0.16 \\
      Phase & 62.76 $\pm$ 14.52 & 0.70 $\pm$ 0.15 \\
      MGD & 70.78 $\pm$ 12.22 & \bftab 0.79 $\pm$ 0.12 \\
      IF & \bftab 72.64 $\pm$ 13.37 & \bftab 0.79 $\pm$ 0.13 \\
     \bottomrule 
   \end{tabularx}
 \end{center}
 \vspace{-0.4cm}
\end{table}

To investigate whether phase representations provide complementary cues to the magnitude representation, we analyze the performance of the dual input CNN operating on the magnitude and different phase representations.
Table~\ref{tbl: perf_dual} presents the performance of the dual input CNN operating on the magnitude and phase spectra, the magnitude and MGD spectra, and the magnitude and IF spectra.
In addition, the performance of the dual input CNN from~\cite{Kodrasi_SPL_2021} operating on the temporal envelope and fine structure representations is also presented.
When comparing the dual input CNNs operating on different magnitude and phase representations, it can be observed that using the magnitude and IF spectra yields the highest performance, with an accuracy of $93.69$\% and an AUC score of $0.97$.
Further, it can be observed that combining the magnitude and phase spectra yields a considerably better performance than combining the magnitude and MGD spectra.
This result shows that although the phase spectrum is irregular, it contains more complementary cues to the magnitude spectrum for dysarthric speech detection than the MGD spectrum.
A comparison of the results in Tables~\ref{tbl: perf_single} and~\ref{tbl: perf_dual} shows that all phase representations contain complementary cues to the magnitude spectrum, with all dual input CNNs yielding a considerably better performance than their single input counterparts.
Finally, Table~\ref{tbl: perf_dual} shows that using the temporal envelope and fine structure representations yields a similar performance as using the magnitude and unprocessed phase representations, but a considerably worse performance than using the magnitude and IF representations.
These results confirm that the performance improvement we obtained in~\cite{Kodrasi_SPL_2021} can be attributed to the incorporation of the analytical phase of the signal and not to the use of auditory-inspired filter banks. 
Nevertheless, exploring alternative representations of the temporal fine structure signals that are applicable for dysarthric speech detection remains a viable future research direction.

In summary, the results presented in this section confirm that all phase representations of the STFT provide useful cues for dysarthric speech detection and should be used in addition to the traditionally used magnitude representation.
In particular, combining the magnitude and IF spectra results in a very high dysarthric speech detection performance.


\begin{table}[t]
  \footnotesize
  \begin{center}
    \caption{\footnotesize Performance of the dual input CNN operating on the magnitude spectrum and different phase representations. The performance of the dual input CNN from~\cite{Kodrasi_SPL_2021} operating on the temporal envelope and fine structure signals is also presented.}
    \label{tbl: perf_dual}
    \begin{tabularx}{\linewidth}{Xrr}
      \toprule
      Representation & Accuracy & AUC \\
      \toprule
      Magnitude-Phase & 87.32 $\pm$ {\color{white}{0}}9.69 & 0.93 $\pm$ 0.10 \\
      Magnitude-MGD & 80.92 $\pm$ 10.11 & 0.90 $\pm$ 0.10 \\
      Magnitude-IF & \bftab 93.68 $\pm$ {\color{white}{0}}5.32 & \bftab 0.97 $\pm$ 0.05 \\
      \hline
      Envelope-Fine structure & 86.04 $\pm$ {\color{white}{0}}8.03 & 0.94 $\pm$ 0.08 \\
     \bottomrule 
   \end{tabularx}
 \end{center}
 \vspace{-0.4cm}
\end{table}

\section{Conclusion}
\label{sec: conc}
Deep learning-based dysarthric speech detection approaches typically learn discriminative representations by processing the magnitude spectrum of signals and ignoring the phase spectrum.
In this paper we have investigated the applicability of STFT phase representations for dysarthric speech detection.
Since the phase spectrum is irregular and visually void of spectro-temporal patterns, we have analyzed two alternative representations which reveal hidden structures of the phase spectrum, i.e., the MGD and IF spectra.
Using a single input CNN we have shown that all considered phase representations, i.e., the unprocessed phase, MGD, and IF spectra, contain dysarthric cues.
Using a dual input CNN operating on both the magnitude and phase representations we have shown that all considered phase representations serve as complementary features to the magnitude spectrum, with the combination of magnitude and IF spectra yielding a high performance.
The presented results have demonstrated the importance of considering phase information for dysarthric speech detection and will hopefully motivate research on novel architectures to optimally combine the magnitude and phase information.

\footnotesize
\bibliographystyle{IEEEbib}
\bibliography{refs}

\begin{thebibliography}{10}

\bibitem{Duffy_book_2003}
J.~R. Duffy,
\newblock {\em {Motor speech disorders: substrates, differential diagnosis, and
  management}},
\newblock Elsevier Mosby, Missouri, USA, 2003.

\bibitem{Rusz_JASA_2013}
J.~Rusz, R.~Cmejla, T.~Tykalova, H.~Ruzickova, J.~Klempir, V.~Majerova,
  J.~Picmausova, J.~Roth, and E.~Ruzicka,
\newblock ``{Imprecise vowel articulation as a potential early marker of
  Parkinson's disease: Effect of speaking task},''
\newblock {\em Journal of the Acoustical Society of America}, vol. 134, no. 3,
  pp. 2171--2181, Sept. 2013.

\bibitem{tracy_bi_2020}
J.~M. Tracy, Y.~{\"O}zkanca, D.~C. Atkins, and R.~H. Ghomi,
\newblock ``{Investigating voice as a biomarker: Deep phenotyping methods for
  early detection of Parkinson's disease},''
\newblock {\em Journal of Biomedical Informatics}, vol. 104, Apr. 2020.

\bibitem{Tsanas_ITBE_2012}
A.~Tsanas, M.~A. Little, P.~E. McSharry, J.~Spielman, and L.~O. Ramig,
\newblock ``{Novel speech signal processing algorithms for high-accuracy
  classification of Parkinson's disease},''
\newblock {\em IEEE Transactions on Biomedical Engineering}, vol. 59, no. 5,
  pp. 1264--1271, May 2012.

\bibitem{Orozco-Arroyave_Interspeech_2015}
J.~R. Orozco-Arroyave, F.~H{\"{o}}nig, J.~Arias-Londo{\~n}o, J.~Bonilla,
  S.~Skodda, J.~Rusz, and E.~N{\"{o}}th,
\newblock ``{Voiced/unvoiced transitions in speech as a potential bio-marker to
  detect Parkinson's disease},''
\newblock in {\em Proc. Annual Conference of the International Speech
  Communication Association}, Dresden, Germany, Sept. 2015, pp. 95--99.

\bibitem{Kodrasi_ITASLP_2019a}
I.~Kodrasi and H.~Bourlard,
\newblock ``Spectro-temporal sparsity characterization for dysarthric speech
  detection,''
\newblock {\em {IEEE} Transactions on Audio, Speech, and Language Processing},
  vol. 28, no. 1, pp. 1210--1222, Apr. 2020.

\bibitem{Janbakhshi_SPL_2020}
P.~Janbakhshi, I.~Kodrasi, and H.~Bourlard,
\newblock ``Subspace-based learning for automatic dysarthric speech
  detection,''
\newblock {\em IEEE Signal Processing Letters}, vol. 28, pp. 96--100, Dec.
  2020.

\bibitem{Hernandez_IS_2020}
A.~Hernandez, E.~J. Yeo, S.~Kim, and M.~Chung,
\newblock ``Dysarthria detection and severity assessment using rhythm-based
  metrics,''
\newblock in {\em Proc. Annual Conference of the International Speech
  Communication Association}, Shanghai, China, Sept. 2020, pp. 2897--2901.

\bibitem{Janbakhshi_ICASSP_2021}
P.~Janbakhshi, I.~Kodrasi, and H.~Bourlard,
\newblock ``{Automatic dysarthric speech detection exploiting pairwise
  distance-based convolutional neural networks},''
\newblock in {\em Proc. IEEE International Conference on Acoustics, Speech, and
  Signal Processing}, Toronto, Canada, May 2021, pp. 7328--7332.

\bibitem{Vasquez2017}
J.~Vasquez, J.~R. Orozco, and E.~Noeth,
\newblock ``Convolutional neural network to model articulation impairments in
  patients with {P}arkinson's disease,''
\newblock in {\em Proc. Annual Conference of the International Speech
  Communication Association}, Stockholm, Sweden, Aug. 2017, pp. 314--318.

\bibitem{Vaiciukyna_SOTSG_2017}
E.~Vaiciukynas, A.~Gelzinis, A.~Verikas, and M.~Bacauskiene,
\newblock ``Parkinson's disease detection from speech using convolutional
  neural networks,''
\newblock in {\em Proc. International Conference on Smart Objects and
  Technologies for Social Good}, Pisa, Italy, Nov. 2017, pp. 206--215.

\bibitem{Kwanghoon_IS_2018}
K.~An, M.~Kim, K.~Teplansky, J.~Green, T.~Campbell, Y.~Yunusova, D.~Heitzman,
  and J.~Wang,
\newblock ``Automatic early detection of {Amyotrophic Lateral Sclerosis} from
  intelligible speech using convolutional neural networks,''
\newblock in {\em Proc. Annual Conference of the International Speech
  Communication Association}, Hyderabad, India, Sept. 2018, pp. 1913--1917.

\bibitem{Vasquez_SC_2020}
J.~C. Vasquez-Correa, T.~Arias-Vergara, M.~Schuster, J.~R. Orozco-Arroyave, and
  E.~Noeth,
\newblock ``Parallel representation learning for the classification of
  pathological speech: {S}tudies on {P}arkinson's disease and cleft lip and
  palate,''
\newblock {\em Speech Communication}, vol. 122, pp. 56--67, Sept. 2020.

\bibitem{Janbakhshi_ITG_2021}
P.~Janbakhshi and I.~Kodrasi,
\newblock ``{Supervised speech representation learning for Parkinson's disease
  classification},''
\newblock in {\em Proc. ITG conference on Speech Communication}, Kiel, Germany,
  Sept. 2021, pp. 154--158.

\bibitem{Mowlaee_INTERSPEECH_2014}
P.~Mowlaee, R.~Saeidi, and Y.~Stylianou,
\newblock ``{INTERSPEECH 2014 Special Session: Phase importance in speech
  processing applications},''
\newblock in {\em Proc. Annual Conference of the International Speech
  Communication Association}, Singapore, Sept. 2014, pp. 1623--1627.

\bibitem{Gerkmann_SPLmag_2015}
T.~Gerkmann, M.~Krawczyk-Becker, and J.~Le Roux,
\newblock ``{Phase processing for single-channel speech enhancement: History
  and recent advances},''
\newblock {\em IEEE Signal Processing Magazine}, vol. 32, no. 2, pp. 55--66,
  Mar. 2015.

\bibitem{Yegnanarayana_ICSLP_1990}
B.~Yegnanarayana, H.~A. Murthy, and V.~R. Ramachandran,
\newblock ``Speech enhancement using group delay functions,''
\newblock in {\em Proc. International Conference on Spoken Language
  Processing}, Kobe, Japan, Nov. 1990, pp. 301--304.

\bibitem{Murthy_ICASSP_2003}
H.~A. Murthy and V.~Gadde,
\newblock ``The modified group delay function and its application to phoneme
  recognition,''
\newblock in {\em Proc. IEEE International Conference on Acoustics, Speech, and
  Signal Processing}, Hong Kong, Apr. 2003, pp. 68--71.

\bibitem{Friedman_ICASSP_1985}
D.~Friedman,
\newblock ``Instantaneous-frequency distribution vs. time: An interpretation of
  the phase structure of speech,''
\newblock in {\em Proc. IEEE International Conference on Acoustics, Speech, and
  Signal Processing}, Florida, USA, Mar. 1985, pp. 1121--1124.

\bibitem{Boashash_IEEE_1992}
B.~Boashash,
\newblock ``{Estimating and interpreting the instantaneous frequency of a
  signal -- Part 1: Fundamentals},''
\newblock {\em Proceedings of the IEEE}, vol. 80, no. 4, pp. 520--538, May
  1992.

\bibitem{Oppenheim_IEEE_1981}
A.~V. Oppenheim J.~S. Lim,
\newblock ``The importance of phase in signals,''
\newblock {\em Proceedings of the IEEE}, vol. 69, no. 5, pp. 529--541, May
  1981.

\bibitem{Wang_ITASLP_1982}
D.~Wang and J.~Lim,
\newblock ``The unimportance of phase in speech enhancement,''
\newblock {\em IEEE Transactions on Acoustics, Speech, and Signal Processing},
  vol. 30, no. 4, pp. 679--681, Aug. 1982.

\bibitem{Gerkmann_SPL_2013}
T.~Gerkmann and M.~Krawczyk,
\newblock ``{MMSE-optimal spectral amplitude estimation given the
  STFT-phase},''
\newblock {\em IEEE Signal Processing Letters}, vol. 20, no. 2, pp. 129--132,
  Feb. 2013.

\bibitem{Hedde_ITASLP_2007}
R.~M. Hegde, H.~A. Murthy, and V.~R.~R. Gadde,
\newblock ``Significance of the modified group delay feature in speech
  recognition,''
\newblock {\em IEEE Transactions on Audio, Speech, and Language Processing},
  vol. 15, no. 1, pp. 190--202, Jan. 2007.

\bibitem{Sehgal_STL_2018}
S.~Sehgal, S.~Cunningham, and P.~Green,
\newblock ``Phase-based feature representations for improving recognition of
  dysarthric speech,''
\newblock in {\em Proc. IEEE Spoken Language Technology Workshop}, Athens,
  Greece, Dec. 2018, pp. 13--20.

\bibitem{Degottex_ITASLP_2011}
G.~Degottex, A.~Roebel, and X.~Rodet,
\newblock ``Phase minimization for glottal model estimation,''
\newblock {\em IEEE Transactions on Audio, Speech, and Language Processing},
  vol. 19, no. 5, pp. 1080--1090, July 2011.

\bibitem{Nakagawa_ITASLP_2012}
S.~Nakagawa, L.~Wang, and S.~Ohtsuka,
\newblock ``{Speaker identification and verification by combining MFCC and
  phase information},''
\newblock {\em IEEE Transactions on Audio, Speech and Language Processing},
  vol. 20, no. 4, pp. 1085--1095, May 2012.

\bibitem{Rajan_INTERSPEECH_2013}
P.~Rajan, T.~Kinnunen, C.~Hanilci, J.~Pohjalainen, and P.~Alku,
\newblock ``Using group delay functions from all-pole models for speaker
  recognition,''
\newblock in {\em Proc. Annual Conference of the International Speech
  Communication Association}, Lyon, France, Aug. 2013, pp. 2489--2493.

\bibitem{Deng_IEEEAcess_2016}
J.~Deng, X.~Xu, Z.~Zhang, S.~Fruehholz, and B.~Schuller,
\newblock ``Exploitation of phase-based features for whispered speech emotion
  recognition,''
\newblock {\em IEEE Access}, vol. 4, pp. 4299--4309, July 2016.

\bibitem{Guo2018}
L.~Guo, L.~Wang, J.~Dang, L.~Zhang, H.~Guan, and X.~Li,
\newblock ``Speech emotion recognition by combining amplitude and phase
  information using convolutional neural network,''
\newblock in {\em Proc. Annual Conference of the International Speech
  Communication Association}, Hyderabad, India, Sept. 2018, pp. 1611--1615.

\bibitem{Kodrasi_SPL_2021}
I.~Kodrasi,
\newblock ``{Temporal envelope and fine structure cues for dysarthric speech
  detection using CNNs},''
\newblock {\em IEEE Signal Processing Letters}, vol. 28, pp. 1853--1857, Aug.
  2021.

\bibitem{Kay_ITASSP_1989}
S.~Kay,
\newblock ``A fast and accurate single frequency estimator,''
\newblock {\em IEEE Transactions on Acoustics, Speech, and Signal Processing},
  vol. 37, no. 12, pp. 1987--1990, Dec. 1989.

\bibitem{Stark_Interspeech_2008}
A.~Stark and K.~Paliwal,
\newblock ``Speech analysis using instantaneous frequency deviation,''
\newblock in {\em Proc. Annual Conference of the International Speech
  Communication Association}, Brisbane, Australia, Jan. 2008, pp. 2602--2605.

\bibitem{Smith_nature_2002}
Z.~M. Smith, B.~Delgutte, and A.~J. Oxenham,
\newblock ``{Chimaeric sounds reveal dichotomies in auditory perception},''
\newblock {\em Nature}, vol. 416, no. 6876, pp. 87--90, Mar. 2002.

\bibitem{GITA}
J.~R. Orozco, J.~D. Arias-Londo{\~n}o, J.~Vargas-Bonilla,
  M.~Gonz{\'a}lez-R{\'a}tiva, and E.~Noeth,
\newblock ``New {S}panish speech corpus database for the analysis of people
  suffering from {P}arkinson's disease,''
\newblock in {\em Proc. International Conference on Language Resources and
  Evaluation}, Reykjavik, Iceland, May 2014, pp. 342--347.

\end{thebibliography}

\end{document}